\documentclass[%
 reprint,
 amsmath,amssymb,
 aps,
]{revtex4-2}

\usepackage{dsfont}
\usepackage{graphicx}
\usepackage{dcolumn}
\usepackage{bm}

\begin{document}
\preprint{APS/123-QED}
\title{A violation of the Tsirelson bound in the pre-quantum theory of trace dynamics}

\author{Rabsan G. Ahmed}
 \email{ms20024@iisermohali.ac.in}
\affiliation{ Indian Institute of Science Education \& Research (IISER) Mohali,\\
Sector 81 SAS Nagar, Manauli, PO 140306 Punjab, India}

\author{Tejinder P. Singh}
 \email{tpsingh@tifr.res.in \quad Address from January 1, 2023 : IUCAA, Pune}
\affiliation{Tata Institute of Fundamental Research,\\
Homi Bhabha Road, Mumbai 400005, India}

\begin{abstract}
\noindent The term Bell's theorem refers to a set of closely related results which imply that quantum mechanics is incompatible with local hidden variable theories. Bell's inequality is the statement that if measurements are performed independently on two space-like separated particles of an entangled pair, the assumption that outcomes depend on hidden variables implies an upper bound on the correlations between the outcomes. Quantum mechanics predicts correlations which violate this upper bound. The CHSH inequality is a specific Bell inequality in which classical correlation (i.e. if local hidden variables exist) can take the maximum value of 2. Quantum mechanics violates this bound, allowing for a higher bound on the correlation,  which can take the maximum value $2\sqrt{2}$, known as the Tsirelson bound. Popescu and Rohrlich showed that the assumption of relativistic causality allows for an even higher bound on the CHSH correlation, this value being 4. Why is the bound coming from causality higher than the Tsirelson bound? Are there relativistic causal dynamical theories which violate the Tsirelson bound? In the present paper we answer this question in the affirmative. We show that the pre-quantum theory of trace dynamics, from which quantum theory is emergent as a thermodynamic approximation, permits the CHSH correlation to take values higher than $2\sqrt{2}$. We interpret our findings to suggest that quantum theory is approximate, and emergent from the more general theory of trace dynamics.

\end{abstract}

\maketitle

\section{\label{sec:1}Introduction}
Correlation between subsystems of a composite system has been of great importance in physics. It has been instrumental in resolving the Einstein-Podolsky-Rosen paradox \cite{Einstein:1935rr} in favour of quantum mechanics. In this resolution, following Bell's theorem using a spin singlet system \cite{Bell:1964kc}, an important quantity was defined, that we will refer to as ``CHSH expression" $F$. Bell's argument implied that for a local hidden variable theory, leading to a classical correlation, there is an upper bound of $2$ on the CHSH expression, known as the ``CHSH inequality" \cite{clauser_proposed_1969} ($F_{\text{classical}} \leq 2$). Subsequently,  Tsirelson \cite{cirelson_quantum_1980} showed that a quantum correlation puts an upper bound of $2\sqrt{2}$ on this expression ($F_{\text{quantum}}\leq 2\sqrt{2}$). Treating  nonlocality, rather than indeterminism, as the axiom, Popescu and Rohrlich \cite{popescu_quantum_1994} demonstrated that for any theory that preserves relativistic causality, the  upper bound on the CHSH expression is $F_{causal}=4$. In that paper the authors speculated the existence of a \textit{supraquantum} theory that can  violate the Tsirelson bound and can account for such a gap between $F_{quantum}$ and $F_{causal}$.

In the present paper, we investigate the CHSH expression in the context of a pre-quantum theory, namely ``Trace Dynamics". Trace dynamics is a matrix valued Lagrangian dynamics, developed by Adler and collaborators, from which quantum field theory can be shown to emerge  as an approximation. This dynamics also has a fundamental conserved charge associated with the system, absent in classical dynamics, from which Heisenberg commutation relations can be shown to emerge. We have briefly summarised the theory in Appendix \ref{A}; however the reader is invited to refer to Adler's textbook on this subject \cite{adler_quantum_2004} as well as the original papers \cite{Adler_1994},\cite{Adler_1996}. The key idea is that trace dynamics holds at Planck time resolution, and if one is observing the system at much lower time resolution, the approximate dynamics can be arrived at by applying the methods of statistical thermodynamics to the underlying theory; the emergent dynamics is quantum theory.

We show that it is possible to find cases where the Tsirelson bound is violated in trace dynamics, leading to a stronger correlation than in quantum mechanics, thus accounting for the aforementioned gap. To analyse the situation in trace dynamics, from which quantum field dynamics naturally emerges \cite{adler_quantum_2004}, we need to develop a formalism similar to quantum field theory for a spin singlet system, preferably for a simple two-dimensional Pauli spinor. There are possibly many ways to do it. However, we focus on a very simple and reasonable case.

The paper is organised as follows. In Sec. \ref{sec:2}, we develop the formalism of spin systems suitable for trace dynamics and show its equivalence to the Pauli algebra. Afterwards spin is considered in the context of trace dynamics in Sec. \ref{sec: 3}. In this section, we derive, the CHSH expression in trace dynamics, $F_{TD}$, and point out  the possibility of violation of the Tsirelson bound. We also present a particular case to explicitly demonstrate the violation, and discuss the case further.

\section{\label{sec:2}Discussion of spin in terms of creation and annihilation operators}
Let us consider a two dimensional column matrix, that transforms as a two-dimensional spinor under rotation in space, with complex numbers as entries:
\begin{equation}
    q=\begin{pmatrix}
    q_1\\q_2
    \end{pmatrix}\nonumber
\end{equation}
We choose a Lagrangian, $L=\Dot{q}^{\dagger}\Dot{q} - q^{\dagger}q = \sum_i \Dot{q_i}^*\Dot{q_i} - {q_i}^*q_i$ and we treat $q_i, q_i^*$ as the generalised coordinates. This immediately gives the equation of motion: $\ddot{q}_i+q_i=0$. We write the general solution as:
\begin{align}
    q_i = \frac{1}{\sqrt{2}}(a_ie^{-i t}+b_i^*e^{i t})
\end{align}
\subsection{Quantization: Spin algebra}
We quantize this theory by promoting $q_i, q_i^{\dagger}$ and their conjugate momenta to operators and specify a commutation relation between them. If we choose the Heisenberg algebra $[q,p]=i$ (We work in $\hbar = 1$ unit), we get $[a_i,a_j^{\dagger}]=[b_i, b_j^{\dagger}]=1$ and all other commutators vanish.

We can also define a vacuum state $|0\rangle$, annihilated by $a_i, b_i$. The normal ordered Hamiltonian, $H$, given by $\sum_{i}(a_i^{\dagger}a_i+b_i^{\dagger}b_i)$, vanishes for the vacuum state. Note, that we primarily get four different kinds of excited states upon applying the creation operators. Moreover, if we define the following three operators, whose correspondence to the conventional spin angular momentum will be demonstrated shortly:
\begin{align}
    S_1^A = \frac{1}{2}(a_1^{\dagger}a_2 + a_2^{\dagger}a_1)\label{1}\\
    S_2^A = \frac{i}{2}(a_2^{\dagger}a_1 - a_1^{\dagger}a_2)\label{2}\\
    S_3^A = \frac{1}{2}(a_1^{\dagger}a_1 - a_2^{\dagger}a_2)\label{3}
\end{align}
they satisfy $[S_i, S_j]=i\epsilon_{ijk}S_k$. Similarly, $S_i^B$ can be defined and they commute with each of the $S_i^A$. All these three operators can be verified to commute with the Hamiltonian, therefore they are constants of motion. Furthermore, $a_i^{\dagger}|0\rangle$ are eigenstates of $S_3^A$ with eigenvalues $+1/2$ and $-1/2$ respectively. Thus one recovers the Pauli algebra of spin-1/2 particles in this formalism. Obviously $a = (a_1, a_2)^T$ transforms like a two dimensional spinor under rotation. Hence $\{a_i^{\dagger}|0\rangle\}$ forms the basis of a faithful representation of $SU(2)$ generated by $\{S_i^A\}$. 

Note that $S_i^{A}$ are just generators of a $SU(2)$ algebra regarding the particle; it does not necessarily have to be the physical spin angular momentum that couples to magnetic field. As this paper only deals with the correlation between two parts of a composite system, it suffices to generally show the results, which is what we have done. In the fermionic case that is discussed in Appendix \ref{C}, these operators have their natural attributes of physical spin angular momentum \cite{Peskin:1995ev}. We will see that the results in this paper equally hold for the fermionic case, justifying our explanation using the much simpler bosonic case.

As $[S_i^A,S_j^B]=0$, for all $i,j$, we can find a natural representation of a two particle system in this formalism. Note $b_i^{\dagger}|0\rangle$ are eigenstates of $S_3^B$ with eigenvalues $+1/2$ and $-1/2$ respectively. Therefore, we interpret $a_1^{\dagger}|0\rangle$ as a state with one particle of $a$-kind with spin up, and $b_1^{\dagger}|0\rangle$ as a state with one particle of $b$-kind with spin up. Interpreting $S_i = S_i^A + S_i^B$, as the $i$-th component of the total spin angular momentum, we get $S_3(a_1^{\dagger}b_1^{\dagger}|0\rangle) = a_1^{\dagger}b_1^{\dagger}|0\rangle$. Therefore, we can say that $a_1^{\dagger}b_1^{\dagger}|0\rangle$ is a state with two particles with spin up in $Z$-direction. One can immediately check that $\{a_1^{\dagger}b_1^{\dagger}|0\rangle, \frac{1}{\sqrt{2}}(a_1^{\dagger}b_2^{\dagger}+a_2^{\dagger}b_1^{\dagger})|0\rangle, a_2^{\dagger}b_2^{\dagger}|0\rangle\}$ forms a triplet and $\frac{1}{\sqrt{2}}(a_1^{\dagger}b_2^{\dagger}-a_2^{\dagger}b_1^{\dagger})|0\rangle$ forms a singlet under the algebra generated by $\{S_i\}$. Our analysis of CHSH in this paper will be focused on this spin singlet state:
\begin{align}
    \boxed {|\Psi\rangle = \frac{1}{\sqrt{2}}(a_1^{\dagger}b_2^{\dagger}-a_2^{\dagger}b_1^{\dagger})|0\rangle}
\end{align}
\subsection{Expectation values and CHSH expression}
It is trivial to note that once we have a state $|\psi\rangle$, the expectation value of an operator $S_i$, say, in this state is given by: $\langle\psi|S_i|\psi\rangle$. We can think of the expectation value in terms of Wightman functions of this theory. For example, in case of our spin singlet the expectation $\langle\psi|S_i^A|\psi\rangle$ can be expanded as:
\begin{equation}
   \frac{1}{4} \langle 0|\{(b_2a_1-b_1a_2)(a_1^{\dagger}a_1 - a_2^{\dagger}a_2)(a_1^{\dagger}b_2^{\dagger}-a_2^{\dagger}b_1^{\dagger})\}|0\rangle 
\end{equation}
Note that ${S_i^{A(B)}}$ transforms like a vector under rotation in space. This can be verified from the transformation of $a_i$ and their substitution in the definitions of ${S_i^{A(B)}}$. 

The CHSH expression, $F$, is defined by
\begin{equation}
    F = |E(C,D) - E(C,D') + E(C',D) + E(C',D')|
\end{equation}
where $C$ and $C'$ are two different measurement axes for the first particle and $D$ and $D'$ are the same for the second particle. $E(M,N)$ is the expectation value of the product of the spin operators of the particles along the axes $M$ (for the first particle) and $N$ (for the second particle). It has been shown that for classical correlations, where the expectation values are specified by a statistical distribution of some local hidden variable, $F$ cannot exceed $2$ for a spin singlet system. A similar upper bound of $2\sqrt{2}$ is there on quantum correlations, due to Tsirelson, which can be obtained, as an example, for the case: $C=Z,C'=X; D=\frac{1}{\sqrt{2}}(X+Z), D'=\frac{1}{\sqrt{2}}(X-Z)$. It can be checked that in our formalism, the CHSH expression reduces for this case into 
\begin{align}
    &F_{\text{quantum}}=\nonumber\\&\frac{1}{4\sqrt{2}}|\langle0|(b_2a_1-b_1a_2)[(a_1^{\dagger}a_1-a_2^{\dagger}a_2)(b_1^{\dagger}b_1-b_2^{\dagger}b_2)\nonumber\\&+(a_1^{\dagger}a_2+a_2^{\dagger}a_1)(b_1^{\dagger}b_2+b_2^{\dagger}b_1)](a_1^{\dagger}b_2^{\dagger}-a_2^{\dagger}b_1^{\dagger})|0\rangle|
    \label{qchsh}
\end{align}
Using the commutation algebra one easily finds $F_{\text{quantum}}$ to be $2\sqrt{2}$, which is the maximum value quantum mechanics allows. Thus, the CHSH expression realising the Tsirelson bound is expressed in this formalism.

Completing our discussion about a QFT-like treatment of Pauli spinors, we now address the trace dynamics level picture, from where qauntum theory emerges.

\section{\label{sec: 3}Trace Dynamical picture of spin}
As is the standard procedure of trace dynamics, we promote $q_i$ to some matrix with even dimension. However, we specify that the matrices are bosonic, i.e. the matrix entries are even-grade Grassmann numbers, as we get commutation relations rather than anticommutation relations in the emergent theory. The trace Lagrangian is $\mathbf{L}=\mathbf{Tr}(\sum_i \Dot{q_i}^{\dagger}\Dot{q_i} - {q_i}^{\dagger}q_i)$. The dynamics is immediately recognised to be similar to Klein-Gordon fields \cite{Peskin:1995ev}. Note that this Lagrangian can be thought of as a simplified form of the bosonic part of the Lagrangian that is postulated in an octonion-valued trace dynamical theory, namely Spontaneous Quantum Gravity \cite{Singh:2022lpw}. Also, a similar analysis could have been done for a fermionic valued underlying trace theory (see Appendix \ref{C}).  

The trace Lagrangian leads to the equation of motion: $\ddot{q}_i+q_i=0$, solutions to which can again be written as
\begin{align}\label{decomposition}
    q_i = \frac{1}{\sqrt{2}}(\alpha_ie^{-i t}+\beta_i^{\dagger}e^{i t})
\end{align}
for some matrices $\alpha_i, \beta_i$, with conjugate momenta $p_i=\Dot{q}_i^{\dagger}$. The trace Hamiltonian is given by $\mathbf{H}=\mathbf{Tr}(\sum_i p_i^{\dagger}p_i + {q_i}^{\dagger}q_i)$ which is invariant under a global unitary transformation. Therefore, we have a conserved Noether charge, known as the Adler-Millard charge of this theory, given by
\begin{equation}
\label{AM charge}
    \tilde{C}=\sum_i [q_i,p_i] + [q_i^{\dagger},p_i^{\dagger}]
\end{equation}
The existence of this charge (which is anti-self-adjoint) is at the heart of the relation between trace dynamics and quantum field theory. It implies that different degrees of freedom exchange $[q,p]$, with the commutator for each of them evolving dynamically, yet the sum of the commutators is conserved. This charge, having the dimensions of action, replaces the Heisenberg algebra in trace dynamics.

Assuming trace dynamics to hold at some energy scale not yet probed by experiments, one asks what is the emergent approximate dynamics at lower energy scales.
Following the standard procedure  of trace dynamics, this charge gets equipartitioned upon thermodynamic approximation, leading each term in the charge to be equal to $i_{eff}\hbar$, where $i_{eff}=i\text{ diag}(1,-1,1,-1,...,1,-1)$. The emergent theory only involves the effective projection of the dynamical variables $x$, denoted by $x_{eff}$ \cite{adler_quantum_2004} (see Appendix \ref{A} for the definition). Relating $q_{eff}$ to some ladder operators and defining a vacuum state in the emergent Hilbert space, we recover the full Pauli algebra in terms of creation and annihilation operators, as described in Sec. \ref{sec:2}
\subsection{Expectation values in Trace Dynamics}
To find a suitable definition of expectation values in trace dynamics, we look at the quantity that upon coarse-graining produces the emergent expectation value in quantum mechanics. Our guiding principle is the correspondence between Wightman functions in the QFTs and the ensemble average of an operator polynomial in trace dynamics, sandwiched between the trace dynamical ground state \cite{adler_quantum_2004}:
\begin{align}
\label{eq: wightman}
    \psi_0^{\dagger}\langle S\{x_{eff}\}\rangle_{AV} \psi_0=\langle0|S\{X\}|0\rangle
\end{align}
As shown earlier, the expectation value of $S_3^A$ in a state $\Psi|0\rangle = \sum_iu_ia_i^{\dagger}|0\rangle$ (normalised), for example, is given by:
$\langle0|\Psi^{\dagger}S_3^A\Psi|0\rangle$
Here, the expression sandwiched between the vacuum state is a polynomial in $a_i,a_i^{\dagger}$, equivalently in $q_i,p_i$. Eqn. (\ref{eq: wightman}) motivates us to write the trace dynamics level expression to be similar but with $\alpha_i,\alpha_i^{\dagger}$ as the variables instead. Further justification is given as follows.

A crucial assumption is made in trace dynamics \footnote{Prof. Stephen Adler mentioned this in a personal communication with RGA, stating that it was implicit in the book.} -
\begin{align}
   &\textit{The part of a variable that anticommutes with $i_{eff}$ is}\nonumber\\&\textit{such that terms quadratic or of higher order in them}\nonumber\\&\textit{can be neglected.} \label{assumption}
\end{align}
This assumption is also applied in deriving the emergent Heisenberg equations of motion. However, using this assumption, in parallel to Eqn. (\ref{qchsh}), the CHSH expression in trace dynamics level, can be written as:
\begin{widetext}
\begin{align}
\label{CHSH TD}
F_{TD} = \frac{1}{4\sqrt{2}}|\psi^{\dagger}_0(\beta_2\alpha_1-\beta_1\alpha_2)[(\alpha_1^{\dagger}\alpha_1-\alpha_2^{\dagger}\alpha_2)(\beta_1^{\dagger}\beta_1-\beta_2^{\dagger}\beta_2)\\+(\alpha_1^{\dagger}\alpha_2+\alpha_2^{\dagger}\alpha_1)(\beta_1^{\dagger}\beta_2+\beta_2^{\dagger}\beta_1)]\nonumber(a_1^{\dagger}\beta_2^{\dagger}-\alpha_2^{\dagger}\beta_1^{\dagger})\psi_0|
\end{align}
\end{widetext}
More justification for choosing this as the CHSH expression will be given in the section on ``Relativistic causality" from an analogous point of view with thermodynamics that seems experimentally meaningful as well.

In contrast to quantum mechanics, where these operators satisfy the Heisenberg algebra as we have seen, in trace dynamics no such commutation relations are assumed. This prevents the reduction of $F_{TD}$ into a trivial expression. As a consequence, $F_{TD}$ appears to have a wide range of possibilities - respecting some normalisations, that we will come to later - many violating Tsirelson bound as well. Once coarse graining takes place, we can immediately see the average of the effective projection of (\ref{CHSH TD}) to produce the result $2\sqrt{2}$ by virtue of the emergent Heisenberg algebra. 

Furthermore, we would like to present a case where the violation can be explicitly realised, which seems interesting. Later on we will provide an interpretation to our findings and its implications for the pre-quantum theories comprising of a trace dynamical formulation, such as the Octonion theory \cite{Singh:2022lpw},\cite{spin},\cite{Priyank}.

\subsection{\label{Case}Violation of the Tsirelson bound}
We break $\alpha, \beta$ into two parts: $\alpha_i = \mathrm{a}_i + \mathcal{A}_i; \beta_i = \mathrm{b}_i + \mathcal{B}_i$, where $\mathrm{a}_i,\mathrm{b}_i$ commute and $\mathcal{A}_i,\mathcal{B}_i$ anticommute with $i_{eff}$. From our assumption (\ref{assumption}) it follows that we can ignore terms quadratic in $\mathcal{A}_i,\mathcal{B}_i$. In particular, we show that for the following range of $\mathrm{a}_i,\mathrm{b}_i,\mathcal{A}_i,\mathcal{B}_i$, the upper bound of $2\sqrt{2}$ is violated:
\begin{align}
[\mathrm{a}_i, \mathrm{a}_j^{\dagger}] = \delta_{ij}\mathds{1};& \quad [\mathrm{b}_i, \mathrm{b}_j^{\dagger}] = \delta_{ij}\mathds{1}\label{a1}\\
[\mathrm{a}_j, \mathcal{A}_i^{\dagger}] \neq 0;& \quad[\mathrm{b}_j, \mathcal{B}_i^{\dagger}] \neq 0\label{a2}\\
\mathrm{a}_i\psi_0 = & \mathrm{b}_i\psi_0 = 0\label{a3}
\end{align}
All the other commutators vanish and we take the normalised $\psi_0$: $\psi_0^{\dagger}\psi_0=1$. After some lengthy but straightforward algebraic manipulations, we get that the CHSH expression reduces to:
\begin{align}
    \boxed{F_{TD}=2\sqrt{2} + \frac{1}{\sqrt{2}}\text{Re}(P_0)}\label{chsh 4}
\end{align}
where $P_0 = \psi_0^{\dagger}P\psi_0$ and $P =\sum_{i,j}\mathcal{A}_i\mathrm{a}_j^{\dagger}+ \mathcal{B}_i\mathrm{b}_j^{\dagger}$

One can always find $\mathcal{A}_i,\mathcal{B}_i$ satisfying (\ref{a2}), such that $P_0$ is positive. For example, once we find $\mathrm{a}_i,\mathrm{b}_i,\mathcal{A}_i,\mathcal{B}_i$ satisfying (\ref{a1}), (\ref{a2}), (\ref{a3}), if we get a negative $P_0$, we can immediately get a positive $P_0$ by replacing $\mathcal{A}_i,\mathcal{B}_i$ by $-\mathcal{A}_i,-\mathcal{B}_i$ as they still satisfy the same conditions. Therefore, (\ref{chsh 4}) describes a legitimate \textbf{violation of the Tsirelson bound}. 

Note that $P_{eff} = 0$, as it anticommutes with $i_{eff}$ and for this particular case, we get the $2\sqrt{2}$ back, consistent with our previous discussion.
\subsubsection{Relativistic Causality}
So far we have not discussed probabilities, which is crucial to predict a sensible experimental result. We assume that the expectation values from experiment performed at energies where trace dynamics effects are relevant (at Planck energy for example), can still be thought of as a statistical average, analogous to quantum mechanics, but over a highly restrictive area of the phase space. This can be thought of as breaking down (\ref{eq: wightman}) into average of contributions from different parts of the phase space. The current experimental techniques can detect only the full average, whereas experiments at higher than Planck energy can break down and give results from more restricted areas of the phase space. This is analogous to the discussion of microstates in the context of thermodynamics and statistical mechanics; after all, quantum mechanics emerges as the equilibrium dynamics of the underlying trace dynamics. Thus, we can conceive a definition of an analogous probability measure.

According to the discussion from the previous paragraph, from (\ref{eq: wightman}) we can realise that expressions of the form $\langle0|\phi(a,a^{\dagger})\psi(a,a^{\dagger})|0\rangle$ in quantum mechanics correspond to an ensemble average over the effective projection of $\psi_0^{\dagger}\phi(\alpha,\alpha^{\dagger})\psi(\alpha,\alpha^{\dagger})\psi_0$. The same can be applied to the concept of probability: the experimental probability density in the vicinity of $\alpha$, let us say, is proportional to $\psi_0^{\dagger}\phi(\alpha,\alpha^{\dagger})\psi(\alpha,\alpha^{\dagger})\psi_0$, with the proportionality coefficient supposedly depending on the details of the ensemble. The failure to give a concept of probability for a general case is linked to the lack of details of the measurement process in trace dynamics level. According to trace dynamics, in quantum mechanics we get Born's probability rule from the stochastic Schrodinger equation \cite{adler_quantum_2004}, that essentially contains the measurement process. However, we find that in trace dynamics, quantities of interest, such as expectation values, can always be expressed in the form of $\psi_0^{\dagger}\phi(\alpha,\alpha^{\dagger})\psi(\alpha,\alpha^{\dagger})\psi_0$, following its quantum counterpart, after all rendering our analysis ultimately consistent. Whatever might be the probability of the individual outcomes, the correspondence (\ref{eq: wightman}) re-affirms the expectation value in trace dynamics as we have defined.       

Therefore, with this definition of a probability measure, we investigate the condition on $\alpha, \beta$ that preserves relativistic causality, as defined by Popescu and Rohrlich without referring to a background spacetime \cite{popescu_quantum_1994}. This definition says that for a correlation in an entangled system to be causal, the probability of measuring spin up or down for a particle along an axis is independent of the choice of axis for the measurement of spin for the other particle. For elaborate demonstration of causality, thus defined, for quantum mechanics, specifically carried out in the present formalism, see Appendix \ref{B}. Note that causality in QFTs like Klein-Gordon theory is preserved using a similar reasoning \cite{Peskin:1995ev}.

Therefore, we find that relativistic causality is always preserved in trace dynamics according to our definition of probability. This can be proven using the transformation properties of the matrices under rotation (see Appendix \ref{B} for the proof). Hence, we are in the regime of causality and validity of the CHSH argument. 

Also, if we use the probability measure to normalise the state in the vicinity of $\alpha,\beta$, we must have 
\begin{equation}
    \frac{1}{2}\psi_0^{\dagger}(\beta_2\alpha_1-\beta_1\alpha_2)(\alpha^{\dagger}_1\beta^{\dagger}_2-\alpha^{\dagger}_2\beta^{\dagger}_1)\psi_0 = 1
\end{equation}
For the case we have chosen, this expression reduces to Re$\left(\sum_i\psi_0^{\dagger}\mathcal{A}_i\mathrm{a}_i^{\dagger}\psi_0+\psi_0^{\dagger} \mathcal{B}_i\mathrm{b}_i^{\dagger}\psi_0\right) = 0$. Therefore, we are left with $P_0=\sum_{i\neq j}\psi_0^{\dagger}\mathcal{A}_i\mathrm{a}_j^{\dagger}\psi_0+\psi_0^{\dagger} \mathcal{B}_i\mathrm{b}_j^{\dagger}\psi_0$

Again $\mathrm{a}_i,\mathrm{b}_i$ are bounded by the condition (\ref{a1}) and $\mathcal{A}_i,\mathcal{B}_j$ are small quantities; hence it appears to be the case that the CHSH expression (\ref{chsh 4}) will be bounded well within the PR bound of $4$.

\subsubsection{Discussion on the Adler-Millard Charge and the Hamiltonian}
The Adler-Millard charge of this theory (\ref{AM charge}) can be further calculated using the decomposition (\ref{decomposition}) to give the simple form:
\begin{align}
    \label{AM Charge 1}
    \tilde{C} = i\sum_{k=1,2} ([\alpha_k,\alpha_k^{\dagger}] + [\beta_k,\beta_k^{\dagger}])
\end{align}
In the emergent quantum theory, the effective projections of each of these commutators become equal to $\mathds{1}_{eff} \equiv i_{eff}/i$. It is easy to check from (\ref{CHSH TD}) that with these commutators (and a few more commutators which vanish in the emergent theory), ensemble average of the effective projection of $F_{TD}$ becomes $2\sqrt{2}$. It is also interesting to note that, if we had chosen the commutators in (\ref{a1}) to be proportional to $\mathds{1}_{eff}$ instead of $\mathds{1}$, $F_{TD}$ would be equal to $2\sqrt{2}$, independent of $\mathcal{A}_i, \mathcal{B}_i$. Therefore, we see that around the ensemble average we have a dense region in phase space satisfying the Tsirelson bound. This is perhaps suggestive of the equilibrium nature of the Tsirelson bound, and quantum mechanics in general. 

The discussion above indicates that when the effective projection of each term of the Adler-Millard charge is equal to their quantum counterpart, no violation of the Tsirelson bound is obtained.

Also, note the trace Hamiltonian mentioned in Sec. \ref{sec: 3}, which upon decomposition (\ref{decomposition}) simplifies to
\begin{align}
    \textbf{H} = \frac{1}{2}\textbf{Tr}(\sum_{i=1,2}\{\alpha_k,\alpha_k^{\dagger}\} + \{\beta_k,\beta_k^{\dagger}\})
\end{align}
where curly bracket denotes anticommutator. The Hamiltonian is self-adjoint, hence the trace Hamiltonian is real. This corresponds to the anti-self-adjointness of the Adler-Millard Charge, $\tilde{C}$, as can be explicitly seen from (\ref{AM Charge 1}). It has been demonstrated that self-adjoint fluctuations of the Adler-Millard charge are responsible for state reduction and spontaneous localisation leading to a classical system \cite{adler_quantum_2004}. As $\tilde{C}$ in our present context is purely anti-self-adjoint, we have a stable trace dynamical system.   

\subsection{Interpretation}
On a logarithmic scale, the Tsirelson bound of $2\sqrt{2}$ lies symmetrically in between the classical bound of $2$ and the $PR$ bound of $4$
\begin{equation}
\ln 2\sqrt{2} - \ln\sqrt{2}\qquad   \ln 2\sqrt2\qquad  \ln 2\sqrt2 + \ln\sqrt2
\end{equation}
According to the expression (\ref{chsh 4}), given a value of $F$ in the trace dynamics regime, one can always map it to a value of $F$ in the quantum regime. The Tsirelson bound serves as an attractor to which trace dynamical systems as well as classical systems evolve, in the following sense. Trace dynamics is operational whenever one is examining the system at Planck time resolution scale; this is as if one is examining the microstates of the system, and these permit $F$ to go up to 4. Whereas, the coarse-grained system, the one being observed at lower than Planck time resolution, can have two possibilities. According to trace dynamics, quantum theory arises at thermodynamic equilibrium (maximum entropy states) after coarse-graining. The Tsirelson bound is hence the `equilibrium' emergent value of $F$. On the other hand, the emergent quantum system can be driven away from equilibrium by stochastic fluctuations and this is how classical macroscopic systems arise in trace dynamics; these are far from equilibrium low entropy states. Classical systems then evolve, across the universe, to become black holes which eventually evaporate by Hawking radiation to return to the thermodynmic equilibrium state, i.e. the one described by quantum theory and hence saturating the Tsirelson bound. Our findings in this paper hence ascribe a physical reality to supraquantum correlations: these will arise if one could observe quantum systems at Planck scale resolution; the so-called microstates with respect to which the quantum description is itself a macro description.

A system of correlated elementary particles in flight can be described either by the laws of quantum mechanics, or more precisely, by the laws of trace dynamics. The response of the classical measuring apparatuses to the incoming particles is apparently insensitive to whether we describe the particles by quantum mechanics or by trace dynamics. However, this is only so because the time of arrival is being measured to a precision much less than Planck time. If it were possible to measure this time of arrival to Planck time resolution, one will be able to observe supraquantum correlations. Unfortunately that is way beyond current technology; nonetheless it might be possible to conceive ways to measure this particular quantum gravity effect (i.e. supraquantum correlations) indirectly through some ingenious experimental setups at accessible lower energies.

Lastly, we mention that the status of spin angular momentum in trace dynamics (TD) is different from that in quantum field theory. Since TD is a Lagrangian dynamics, spin angular momentum ought to be the canonical momentum corresponding to some appropriate angle in space-time. Obviously, this angle cannot be in our four dimensional space-time, and this was one amongst several reasons which led us to suggest that trace dynamics requires extra dimensions \cite{spin}, and in particular 8D space labelled by the octonions has been suggested because of its promise for a theory of unification \cite{Priyank}. Spin angular momentum, though conserved, is not quantized in units of $\hbar$ unless and until quantum theory emerges in the coarse-grained approximation. Thus in our analysis above, it is implicitly understood that as long as the Adler-Millard charge is not equipartitioned, the spin statistics connection does not hold. This however in no way takes away our conclusions relating to the violation of the Tsirelson bound in trace dynamics, because bosonic and fermionic degrees of freedom in TD are defined not by spin, but by their associated Grassmann-number valued matrices: odd-grade Grassmann for fermions and even-grade Grassmann for bosons. It also follows that in the octonionic theory of unification \cite{Priyank} based on trace dynamics, correlations are supraquantum.

\begin{acknowledgments}
RGA thanks Prof. Stephen L. Adler for his valuable suggestions and insight into trace dynamics.
\end{acknowledgments}

\appendix
\section{\label{A}Trace Dynamics}
The standard procedure of quantum mechanics can be listed out as follows: \textit{Step 1}: write out the classical Hamiltonian of a system in a specific set of canonical variables and promote the canonical variables, and hence all observables that are functions of them, as operators on a Hilbert space comprising of the state vectors of the system. \textit{Step 2}: Specify the commutation relations among the canonical variables, namely Heisenberg algebra $[q,p]=i\hbar$, using correspondence principle. However, although consistent with present experiments, this might seem restrictive, as the canonical variables need to satisfy a commutation relation that is given by the dynamics itself (for example, satisfaction of the Poisson bracket is given by the classical dynamics itself). 

This leads one to the consideration of a theory where the Step 1, which is quite reasonable, is followed. However, one does not specify a commutation relation between the variables, instead the Heisenberg algebra emerges as an approximation. Trace Dynamics is such a theory. We summarise this theory in this appendix and completely refer to the monograph for this topic \cite{adler_quantum_2004} throughout.

In this theory, we take the Lagrangian, $L(q_i,\Dot{q_i})$ of a system and raise $q_i,\Dot{q}_i$ to matrix valued functions with complex Grassmann numbers of bosonic or fermionic kind as entries and define a Trace Lagrangian, $\mathbf{L} = \mathbf{Tr} L(q_i,\Dot{q}_i))$. From the principle of least action, one obtains the Euler-Lagrange equation of motion in Trace Dynamics \cite{adler_quantum_2004}:
\begin{align}
    \frac{d}{dt}\left(\frac{\delta \mathbf{L}}{\delta \Dot{q}_i}\right) = \frac{\delta \mathbf{L}}{\delta q_i}
\end{align}
For the definition of trace derivatives, see \cite{adler_quantum_2004}. One can proceed to find the Trace Hamiltonian from here: $\mathbf{H}=\mathbf{Tr}(\sum_i p_i\Dot{q}_i-L)$, and also derive matrix valued Hamilton's equations of motion. These are in place of the Heisenberg equations of motion of quantum theory; the latter are emergent from the former.
\subsection{Adler-Millard Charge}
If the Trace Hamiltonian in invariant under a global unitary transformation: $q_i \rightarrow U^{\dagger}q_iU$, $p_i \rightarrow U^{\dagger}p_iU$, one can show the existence of a conserved charge \cite{adler_quantum_2004}, namely the Adler-Millard charge, given by
\begin{align}
    \tilde{C} = \sum_{i\in B} [q_i,p_i] - \sum_{i \in F} \{q_i,p_i\} \label{charge}
\end{align}
Here $B(F)$ stands for the set of bosonic (fermionic) variables. Upon coarse-graining, we see this charge getting equipartitioned to give rise to Heisenberg commutation relations in the emergent theory that is Quantum Mechanics. Thus we recover quantum mechanics from a more generalised theory, trace dynamics.
\subsection{Coarse-Graining: Emergence of Heisenberg Algebra and Heisenberg equations of motion}
To apply the methods of statistical mechanics, a measure is defined on the phase space (The set of all canonical variables, $\{q_i,p_i\}\equiv x_r$ and $(x_r)_{mn}=(x_r)_{mn}^0+i(x_r)_{mn}^1$)
\begin{align}
    d\mu = \prod_A d\mu^A\\
    d\mu^A = \prod_{r,m,n} d(x_r)^A_{mn}
\end{align}
A canonical ensemble is defined with constraints determined by conserved quantities such as the trace Hamiltonian, Adler-Millard Charge and a quantity called fermion number \cite{adler_quantum_2004}. The anti-self-adjointness of $\tilde{C}$ implies that its canonical average $\langle\tilde{C}\rangle_{AV}$ can be written in the following form
\begin{align}
        \langle\tilde{C}\rangle_{AV} = i_{eff} \hbar
\end{align}
where, $i_{eff}= i$ diag$(1,-1,1,-1,...,1,-1)$. 
The equilibrium distribution, under the mentioned constraints, is thus given by
\begin{align}
    \rho =Z^{-1} \exp(-\text{Tr}\tilde{\lambda}\tilde{C} - \tau \mathbf{H}-\eta \mathbf{N})\\
    Z = \int d\mu \exp(-\text{Tr}\tilde{\lambda}\tilde{C} - \tau \mathbf{H}-\eta \mathbf{N})
\end{align}
We can derive the generalized Ward identities given a set of assumptions, elaborated in the book, and as special cases, we derive that under ensemble average the following equalities hold
\begin{align}
    &[q_{reff},p_{reff}] = i_{eff}\hbar \text{ for, }r\in B\\
    &\{q_{reff},p_{reff}\} = i_{eff}\hbar \text{ for, }r\in F\\
    &\hbar \Dot{x}_{reff} = i_{eff}[H_{eff},x_{reff}]
\end{align}
Here, effective projection of a matrix $M$, $M_{eff} \equiv \frac{1}{2}(M-i_{eff}Mi_{eff})$ is the part of $M$ that commutes with $i_{eff}$. Note the emergence of Heisenberg commutation relations and Heisenberg dynamics only through the effective projections of the underlying trace dynamical variable. Although we get two sectors, namely $\pm i$ sectors. We mainly work with $+i$ sector in the standard quantum mechanics. The $-i$ sector has been proposed to account for dark matter \cite{adler_quantum_2004}, \cite{adler_shadow_2013}. Interestingly, the assumption (\ref{assumption}) corresponds to assuming a weakly coupled dark matter sector \footnotemark[\value{footnote}]. Further considerations from the Ward identities in the case of a polynomial of the canonical variables, lead to the correspondence
\begin{align}
    \psi_0^{\dagger}\langle S\{x_{eff}\}\rangle_{AV} \psi_0=\langle0|S\{X\}|0\rangle
\end{align}
Here $|0\rangle$ is the vacuum state of the emergent theory and $\psi_0$ is the eigenstate of $H_{eff}$ with the lowest eigenvalue.

In chapter 6 of \cite{adler_quantum_2004}, it is demonstrated how the measurement process (collapse of the wave function) can be modelled in terms of a stochastic Schrodinger equation, arising from self-adjoint fluctuations of $\tilde{C}$ around its ensemble average. Born probability law is derived as a result, under the assumption that the stochastic modifications of the Schrodinger equation are norm-preserving and do not permit superluminal signalling. Thus the basic concepts of trace dynamics are outlined.
\section{\label{B}Proof of relativistic causality}
Let us take a state of a two particle system in quantum mechanics, $\phi|0\rangle$. Let observer $A$ measures the first particle, in $Z$-basis and observer $B$ also measures the second particle in $Z$-basis. The probability of $A$ measuring spin-up is then given by:
\begin{align}
    P_{ZZ}(+) = &|\langle 0| b_1a_1\phi|0\rangle|^2 + |\langle 0| b_2a_1\phi|0\rangle|^2\nonumber\\
    &= \langle 0| \phi^{\dagger}a_1^{\dagger}a_1\phi|0\rangle\label{B1}
\end{align}
The last line follows from the completeness of the basis ${b_i^{\dagger}}$ for the two dimensional representation of the algebra generated by $\{S_i^B\}$ as illustrated in Sec. \ref{sec:2}. However, if observer $B$ instead measured the second particle in an arbitrary $\mathcal{Z}$-basis, the probability for the same would have been given by:
\begin{align}
    P_{Z\mathcal{Z}}(+) &= |\langle 0|b_1'a_1\phi|0\rangle|^2 + |\langle 0|b_2'a_1\phi|0\rangle|^2\nonumber\\
    &= \langle 0| \phi^{\dagger}a_1^{\dagger}a_1\phi|0\rangle\label{B2}
\end{align}
The last line is related to the fact that the basis of measurement along $\mathcal{Z}$ is related to the basis of measurement along $Z$ by a unitary, $U\in SU(2)$. Hence, we obtain $P_{ZZ}(+)=P_{Z\mathcal{Z}}(+)$. We could have done the same analysis for either of the observers and  gotten the same result. This indicates that relativistic causality, as shown by Popescu and Rohrlich \cite{popescu_quantum_1994}, is preserved in quantum mechanics. We proceed to demonstrate the same for trace dynamics. As defined in Sec. \ref{sec: 3}, the probability when both observers measure in $Z$-basis, can be written from (\ref{B1}) using the correspondence, as discussed in sec. \ref{sec: 3}:
\begin{align}
    P'_{ZZ}(+) \propto \psi_0^{\dagger}\phi^{\dagger} \alpha_1^{\dagger}\alpha_1\phi\psi_0\label{B3}
\end{align}
The probability for the same when $B$ performs the measurement along some arbitrary direction, $\mathcal{Z}$, $P'_{Z\mathcal{Z}}(+)$ corresponds to (\ref{B2}), which would eventually establish: $P'_{ZZ}(+)=P'_{Z\mathcal{Z}}(+)$, by comparing with (\ref{B3}).

This completes the proof. Note that we could equally do the same thing for the first particle. Therefore, our trace dynamical treatment of spin singlet system is causal.

\section{\label{C}Similar analysis in Fermionic variable}
We could similarly work with fermionic variables and a Lagrangian, $L=iq^{\dagger}\gamma \Dot{q}-q^{\dagger}q$ (Note the similarity of its dynamics to the Dirac field \cite{Peskin:1995ev}). This can be again found in the fermionic sector of the Lagrangian in octonionic theory \cite{Singh:2022lpw}; the constants $\alpha$, $\beta_1$, $\beta_2$ in \cite{Singh:2022lpw} have been speculated to be related to the gamma matrices in Dirac spinor formalism, which is consistent with our discussion. 

However, in this case, we rather take the anti-commutator of the canonical variables, as it emerges from the Adler-Millard charge in the underlying trace dynamics \cite{adler_quantum_2004}. Again using the definitions of spin operators from Sec. \ref{sec:2} and the anti-commutation relations, one can show $[S_i, S_j]=i\epsilon_{ijk}S_k$. In contrast to the bosonic case, where we have just demonstrated the correlation between two sets of $SU(2)$ generators, corresponding to the two particles, here these generators correspond to the physical spin angular momentum of the particles, in the context of quantum field theories \cite{Peskin:1995ev}. However, as discussed above, a more precise and consistent way of defining spin in a trace dynamical formalism requires the introduction of extra dimensions \cite{spin}. 

The discussion about the underlying trace dynamical picture follows similarly to the case of bosonic variables. We get the same expression for $F_{TD}$ as in (\ref{CHSH TD}). As for the case we demonstrated, we replace the commutation relations (\ref{a1}), (\ref{a2}) by anti-commutation relations (all other anticommutators vanish, in this case). It is interesting that we get the same expression for $F_{TD}$ as in (\ref{chsh 4}).

For the sake of completeness of the formalism introduced in Sec. \ref{sec:2}, in bosonic case, $\{a_1^{\dagger}|0\rangle, a_2^{\dagger}|0\rangle\}$ forms a $2$-dimensional representation of $SU(2)$, while $\{a_1^{\dagger}a_1^{\dagger}|0\rangle, a_1^{\dagger}a_2^{\dagger}|0\rangle, a_2^{\dagger}a_2^{\dagger}|0\rangle\}$ forms a $3$-dimensional representation of $SU(2)$, corresponding to a ``spin-1" situation. In general, an $n$-dimensional representation of $SU(2)$ is given by
\begin{align}
    \{(a_1^{\dagger})^l(a_2^{\dagger})^m|0\rangle: l+m = n-1, 0\leq l,m < n\}
\end{align}

\nocite{*}
\bibliography{citations}
\end{document}